\documentclass[12pt]{article}
\usepackage{times}
\usepackage{amsmath}
\usepackage{amsfonts}
\usepackage{amssymb}
\usepackage{graphicx}
\usepackage{bm}
\usepackage[colorlinks=true,citecolor=blue,urlcolor=blue,linkcolor=blue,hyperfigures=true]{hyperref}
\usepackage{multirow}
\usepackage{comment}
\usepackage{physics}
\usepackage{ulem}
\usepackage{braket}
\usepackage{gensymb}
\usepackage[a4paper, margin=1in]{geometry}
\usepackage{siunitx}
\usepackage[font=small,labelfont=bf]{caption}
\usepackage{subcaption}
\usepackage{upgreek}
\usepackage{multirow}
\def\Eq{Eq.~}

\def\be{\begin{equation}}
\def\ee{\end{equation}}
\def\ba{\begin{align}}
\def\ea{\end{align}}

\title{Atom interferometry at arbitrary orientations and rotation rates}

\author
{Quentin d'Armagnac de Castanet$^{1,2}$,
Cyrille Des Cognets$^{2}$,
Romain Arguel$^{2,3}$,\\
Simon Templier$^{1}$,
Vincent Jarlaud$^{1,2,\ast}$,
Vincent Ménoret$^{1,2}$,\\
Bruno Desruelle$^{1}$,
Philippe Bouyer$^{1,4,5,6}$,
and Baptiste Battelier$^{2}$
\\
\normalsize{${}^{1}$Exail, 1 rue François Mitterrand, 33400 Talence, France}\\
\normalsize{${}^{2}$LP2N, Laboratoire Photonique, Numérique et Nanosciences,}\\
\normalsize{Université Bordeaux--IOGS--CNRS:UMR 5298,}\\
\normalsize{1 rue François Mitterrand, 33400 Talence, France}\\
\normalsize{$^3$Centre National d'Etudes Spatiales, 18 avenue Edouard Belin, 31400 Toulouse, France}\\
\normalsize{$^4$Van der Waals-Zeeman Institute, Institute of Physics, University of Amsterdam}\\
\normalsize{Science Park 904, 1098XH Amsterdam, Netherlands}\\
\normalsize{$^5$QuSoft, Science Park 123, 1098XG Amsterdam, Netherlands}\\
\normalsize{$^6$Eindhoven University of Technology, Eindhoven, Netherlands}
\\
\normalsize{$^\ast$Corresponding author. E-mail:  vincent.jarlaud@exail.com}
}

\begin{document}
%%%%%%%%%%%%%%%%%%%%%%%%%%%%%%%%%%%%%%%%%%%%%%%%%%%%%%%%%%%%%%%%%%%%%%%%%%%%

\maketitle

%===========================================================================


\begin{quote}
\bf{The exquisite precision of atom interferometers has sparked the interest of a large community for use cases ranging from fundamental physics to  geodesy and inertial navigation. However, their practical use for onboard applications is still limited, not least because rotation and acceleration are intertwined in a single phase shift in free-fall atom interferometers, which makes the extraction of a useful signal more challenging. Moreover, the spatial separation of the wave packets due to rotations leads to a loss of signal. Here we present an atom interferometer operating over a large range of random angles, rotation rates and accelerations. An accurate model of the expected phase shift allows us to untangle the rotation and acceleration signals. We also implement a real-time compensation system using two fibre-optic gyroscopes and a tip-tilt platform to rotate the reference mirror and maintain the full contrast of the atom interferometer. Using these theoretical and practical tools, we reconstruct the fringes and demonstrate a single-shot sensitivity to acceleration of $24$ $\upmu$g, for a total interrogation time of $2T = \SI{20}{\milli\second}$, for angles and rotation rates reaching $\SI{30}{\degree}$ and $\SI{14}{\degree\per\second}$ respectively. Our hybrid rotating atom interferometer unlocks the full potential of quantum inertial sensors for onboard applications, such as autonomous navigation or gravity mapping.}
\end{quote}

%===========================================================================
\section*{Introduction}
\label{sec:Intro}
Atom interferometers rely on the coherent splitting and recombination of the wave packets associated with an ensemble of laser-cooled atoms. When the atoms are in free fall and manipulated by the field of a retroreflected laser beam, the interferometer becomes highly sensitive to accelerations and rotations \cite{Geiger2020,Kasevich1991,Merlet2010,Gustavson1997,Lenef1997,Dutta2016,Savoie2018,Chen2019}. This led to the development of novel inertial sensors with field \cite{Antoni-Micollier2022}, and onboard implementations \cite{Geiger2011,Bidel2018,Bidel2020}, but further development is required to reach the full potential of atom interferometers and their applications in areas including fundamental physics \cite{Aguilera2014,El-Neaj2020}, geodesy \cite{Leveque2021} and navigation \cite{Jekeli2005}.

Rotations and accelerations both contribute to the phase shift of atom interferometers, and instruments are usually designed in a way that the sensitivity to one or the other is cancelled out. For example, gravimeters are operated with a vertical laser beam such that there is no physical area leading to a rotation sensitivity \cite{Peters2001}, and gyroscopes, using two counter-propagating atomic clouds, have been used to separate rotations from accelerations \cite{Gustavson1997,Canuel2006,Gauguet2009,Muller2009}. When the sole configuration of the experiment is no longer sufficient to separate the effect of rotations and accelerations, these techniques are sometimes extended. This is for example the case in long-baseline atom interferometers, where the rotation of the Earth can result in spurious phase shifts and contrast losses. This effect can be mitigated by applying a rotation to the reference mirror which retroreflects the laser beam \cite{Lan2012, Sugarbaker2013}. However, this separation of acceleration and rotation cannot be implemented in the case of strapdown onboard applications, because it relies on a static operation of the inertial sensor, where the geometry of the matter-wave interferometer is known and constant in time. Even in the architectures where rotations are compensated, this is done using \textit{a priori} knowledge of the rotation rate \cite{Zhao2021, Zahzam2022}. In the case of a moving carrier this is no longer possible, and both accelerations and rotations will play a significant part in the measurement. 

Here we demonstrate a new method to retrieve the rotation and acceleration signals simultaneously and independently while maintaining the full contrast of the atom interferometer, by implementing an hybridization with classical inertial sensors. We report on the operation of an atom interferometer over a large angle range $[0,30]\degree$ and for high rotations rates up to $\SI{250}{\milli\radian\per\second}$ ($\SI{14}{\degree\per\second}$). 

First, we implement a double hybridization of our single-axis atom interferometer with a classical accelerometer and two fibre-optic gyroscopes (FOG) and demonstrate a quantitative understanding of the atomic phase shift due to rotations while the phase shift due to acceleration is cancelled out thanks to a real-time closed-loop servo-lock. This method is validated over an intermediate range of rotation rates where the contrast is not fully cancelled by the spatial separation of the wave packets. Then, for higher rotation rates we apply a real-time compensation of the orientation of the reference mirror, using the FOGs' signals to maintain the effective wavevector of the laser along a constant direction and thus ensure the overlap of the wave packets at the output of the atom interferometer. We extend our model of the expected phase shift to fully reconstruct the atomic fringes of our hybrid quantum sensor. 

%===========================================================================
\section*{Results}

Our experiments are performed on a three-axis hybrid atom interferometer setup, installed on a manual three-axis rotation platform and capable of measuring accelerations along three orthogonal directions as described in \cite{Templier2022}.

An ensemble of rubidium 87 atoms is trapped and cooled down to a temperature of approximately $\SI{4}{\micro\kelvin}$. The atoms are then released and prepared in the magnetically insensitive $\ket{F = 1, m_F = 0}$ sub-level of the $5^2S_{1/2}$ ground state. Following a time of flight $T_{\text{TOF}}$, a sequence of three counter-propagating Raman pulses ($\pi/2-\pi-\pi/2$), each separated by the interrogation time $T$, is carried out to coherently split, reflect and recombine atomic wave packets and perform a Mach-Zehnder interferometer. A state-sensitive fluorescence detection allows us to measure the atom number in the hyperfine level $\ket{F = 2, m_F=0}$ as well as the total atom number, respectively denoted as $N_2$ and $N_{tot}$. The interferometer's output signal is the population ratio $R = N_2/N_{tot}$ at the end of the sequence.

\begin{figure}
    \centering
    \includegraphics[width=1.0\textwidth]{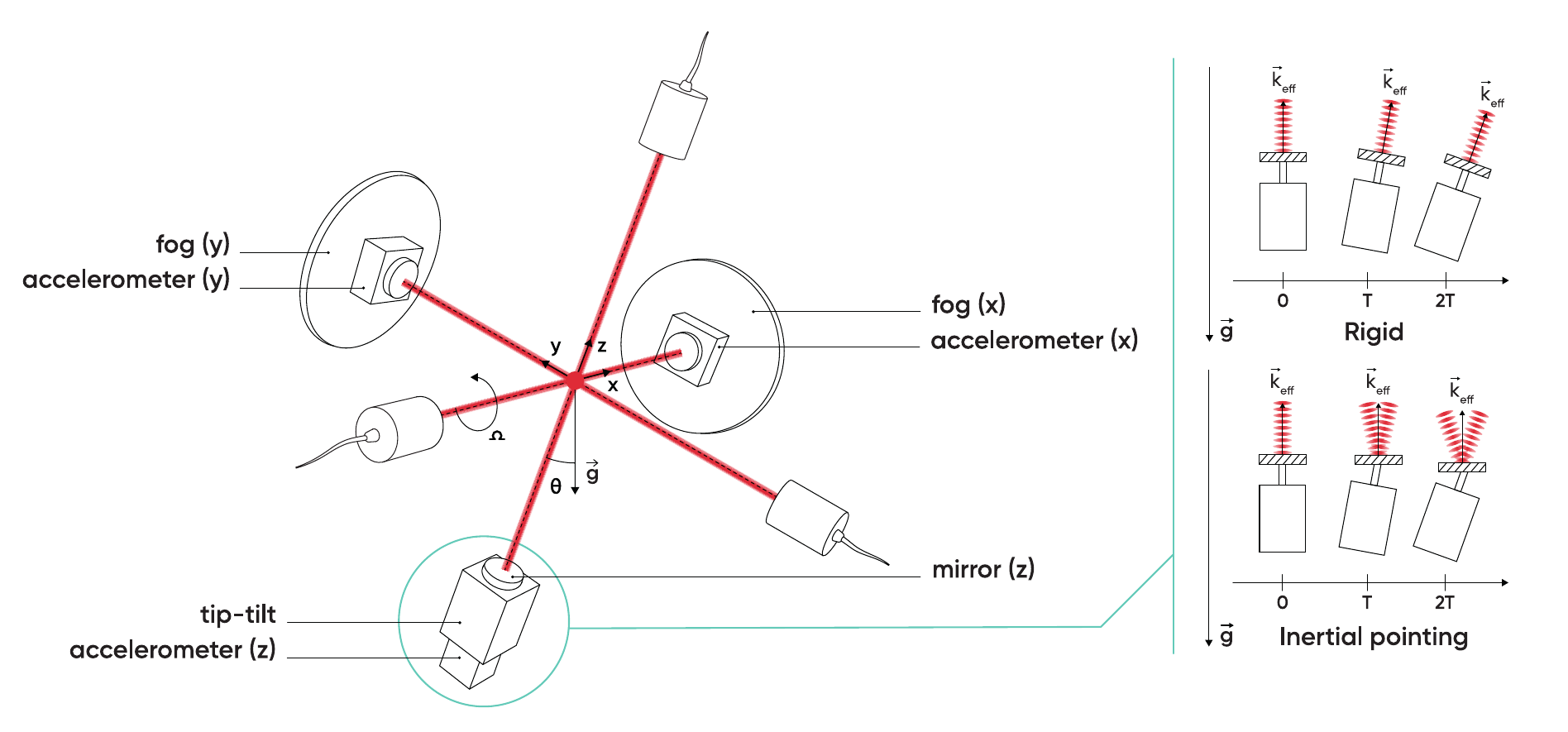}
    \caption{\textbf{Schematic representation of the rotating atom interferometer}. Three orthogonal laser beams are used to trap and cool an ensemble of $^{87}$Rb atoms in a retroreflected configuration. Mechanical accelerometers are attached to the back of the three mirrors and two fibre-optic gyroscopes (FOG) measure the rotations of the device along the $x$ and $y$ axes. The atom interferometer is performed on the $z$ axis, where the reference mirror is mounted on a piezoelectric tip-tilt platform. The experiment rotates with a rotation axis $x$ and a rotation rate $\Omega$. The tilt angle of the experiment is $\theta=\SI{0}{\degree}$ when the axis $z$ is vertical, i.e. aligned with gravity. Inset: the two modes of operation of the tip-tilt mirror represented by a spatio-temporal diagram. In the rigid mode, the tip-tilt behaves like a fixed mirror and follows the rotation of the device. In the inertial pointing mode, the mirror keeps a constant orientation in the frame of the laboratory for the duration of the interferometer ($2T$).}
    \label{fig:manip}
\end{figure}

Here, we perform interferometry along a single axis $z$ and define the reference angle $\theta=\SI{0}{\degree}$ when this axis is vertical, i.e. aligned with gravity. Classical accelerometers attached to the back of the retroreflection mirrors (Fig. \ref{fig:manip}) on each axis provide the signals necessary to calculate the laser phase and frequency shifts required to compensate the tilt-dependant Doppler effect and the noise due to vibrations of the reference mirror \cite{Templier2022}. For all the data presented in this paper, the phase shift due to acceleration is cancelled out in a closed-loop mode as demonstrated in \cite{Templier2022} and described in the Methods. The present study focuses on the phase shift due to rotation.

Two fibre-optic gyroscopes measure the rotation rates around the $x$ and $y$ axes. These FOGs, adapted from a rotational seismometer (Exail --- formerly iXblue --- BlueSeis 3A), are capable of measuring angular velocities up to $\SI{667}{\milli\radian\per\second}$ (approximately $\SI{38}{\degree\per\second}$) with a noise below $\SI{100}{\nano\radian.s^{-1}.Hz^{-1/2}}$.
The device rotates around the $x$ axis, orthogonal to the measurement axis $z$. On the $z$ axis, the reference mirror is mounted on a two-axis piezoelectric tip-tilt platform (Physik Instrumente S-335.2SHM2) in order to dynamically control its orientation. We neglect Earth's rotation which is three to four orders of magnitude below the rotation rate of the device.

\subsection*{Contrast loss and phase shift of a rotating atom interferometer} %%%%%%%%%%%%%%%%%%%%%%%%%%%%%%%%%%%%%%%%%%%%%%%%%%%%%
In this section, the orientation of the mirror remains fixed with respect to the vacuum chamber (rigid mode in Fig. \ref{fig:manip}). The rotation of the device impacts both the contrast and the phase of the Mach-Zehnder atom interferometer, leading to the measured population ratio: 
\begin{align}
   R = R_0 - \frac{1}{2} C(\Omega) \cos{(\phi_r(\Omega))}
    \label{sinusoidOmega}
\end{align}
where $R_0$ is the offset of the interference fringes, $\phi_r(\Omega)$ the phase shift due to the rotation of the experiment and $C(\Omega)$ the contrast measured for a rotation rate $\Omega$.

The rotation of the atom interferometer changes the direction of the laser beam, and therefore, the direction of the momentum transferred to the atoms between two consecutive pulses. This angle leads to a separation in position and momentum of the wave packets associated with the two arms of the interferometer and, as a consequence, to a loss of contrast of the interference fringes \cite{Roura2014}:
\begin{align}
    C(\Omega) \approx C_0\exp\left(-\left(\sqrt{\frac{2 k_B \mathcal{T}}{m}}k_{\text{eff}} T^2 \Omega\right)^2\right)
    \label{contrastLoss}
\end{align}
where $C_0$ is the static contrast of the interferometer i.e. when there is no rotation, $\mathcal{T}$ the temperature of the atomic cloud, $k_B$, the Boltzmann constant, $m$, the mass of the atom and $k_{\text{eff}}$, the effective wavevector of the Raman transition.

The phase term $\phi_r$ in equation \ref{sinusoidOmega} is impacted by the complex trajectory of the free-falling atoms in the reference frame of the rotating laser beam. This leads to phase shifts which are not compensated for by the hybridization with the classical accelerometer. Using the mid-point theorem \cite{Antoine2003a}, we have carried out a derivation of the rotation-induced phase term $\phi_r^{rigid}$ in the rigid mode (see Methods and \cite{Beaufils2023}). Here, we assume the rotation rate $\Omega$ and the angle $\theta$ between the laser beam and the vertical defined by gravity to be constant during the atom interferometer. We also assume that the atoms have no initial velocity when they are released from the optical molasses.

Neglecting the duration of the Raman pulses, we find:
\begin{align}
    \phi_r^{rigid}(\Omega, \theta)=k_{\text{eff}}T^2\left(z_0^{\text{CA}}\Omega^2 - 2\Omega g\sin{(\theta)}(T+T_{\text{TOF}})\right)
    \label{phi_noComp}
\end{align}
where $g$ is the free-fall acceleration of the atoms under the effect of gravity ($g\simeq\SI{-9.806}{\meter\per\second\squared}$), $z_0^{\text{CA}}$ denotes the distance between the atomic wave packet and the classical accelerometer at the beginning of the interferometer, $T_{\text{TOF}}$ the time of flight before the first interferometer pulse. We find $z_0^{\text{CA}} = \SI{204}{\milli\meter}$ for $\theta = \SI{0}{\degree}$ and $T_{\text{TOF}} = \SI{20}{\milli\second}$ (see Methods). The first term in this expression corresponds to a differential centrifugal acceleration between the classical accelerometer and the atom interferometer due to the different positions of their respective test masses. The second term of equation \ref{phi_noComp} is akin to a Coriolis acceleration and depends on the velocity component of the atoms perpendicular to the direction of propagation of the laser. For significant tilts ($\theta \gtrsim \SI{3}{\degree}$), this term is the dominant one with the typical parameters of our experiment. Both terms are factored by the scale factor of the atom interferometer $k_{\text{eff}}T^2$. For numerical applications, the average values of $\theta$ and $\Omega$ during the interferometer are used. Note that this model does not take into account terms associated with a tangential acceleration of the atoms which depends on the angular acceleration of the device (see Methods).

\begin{figure}
    \centering
    \includegraphics[width=\textwidth]{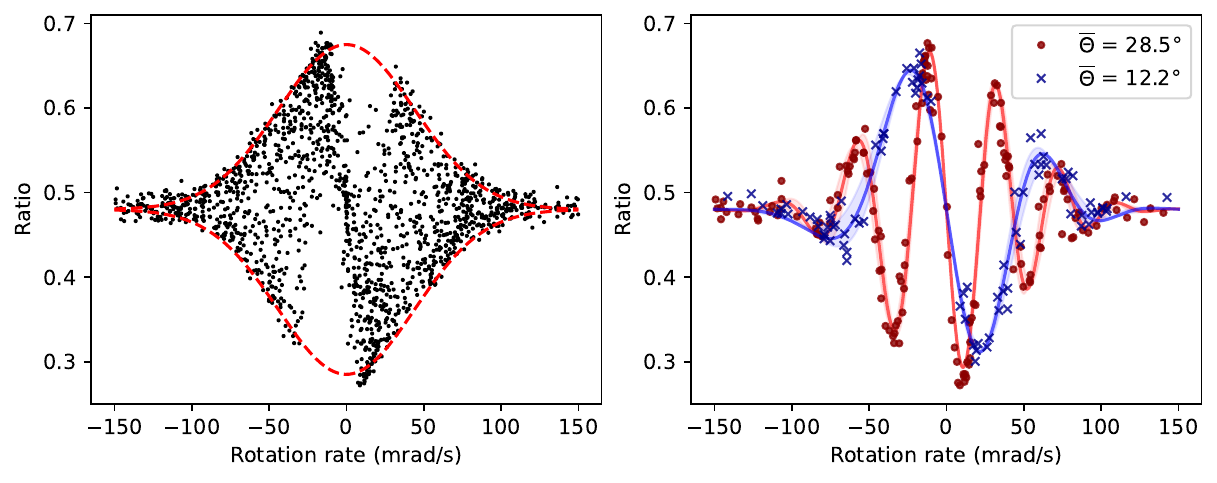}
    \caption{\textbf{Rotating atom interferometer in the rigid mode}. Population ratio $N_2/N_{tot}$ as a function of the rotation rate of the device. \textbf{Left:} 1600 measurements with $\theta \in [0,30]\degree$, $2T=\SI{12}{\milli\second}$. The dashed red line is an envelop defined by $y = R_0\pm C(\Omega)/2$. \textbf{Right:} Output of the atom interferometer for a tilt angle $\theta \in [11,13]\degree$ (blue) and $\theta \in [27,30]\degree$ (red). The solid lines are the theoretical curves calculated using equations \ref{sinusoidOmega}, \ref{contrastLoss} and \ref{phi_noComp}. The shaded regions correspond to the model values for the extrema of the angle ranges.}
    \label{ratioVsRotationRate}
\end{figure}

The data presented in figure \ref{ratioVsRotationRate} (left) show the population ratio $R$ measured at the output of the rotating atom interferometer, as a function of the angular velocity. The result includes approximately 1600 measurement cycles with an interrogation time $2T=\SI{12}{\milli\second}$, angular velocities up to $\SI{150}{\milli\radian\per\second}$ ($\SI{8.6}{\degree\per\second}$) and tilts of the device between $\SI{0}{\degree}$ (vertical) and $+\SI{30}{\degree}$. A constant $\pi/2$ phase offset of the laser is applied on the last pulse of the interferometer which is why $R\approx 0.5$ for $\Omega = 0$. The rotation rate of the device is directly given by the measurement of the gyroscope while the angle $\theta$ is calculated from the accelerometer's signal. The gyroscopes are sampled at $\SI{1}{\kilo\hertz}$ and the accelerometers at $\SI{2.5} {\kilo\hertz}$, allowing us to average out the inertial signals over the duration of the interferometer and reduce the influence of electronic noise. The distribution of the data points highlights the loss of contrast for increasing angular velocities. The points are contained in an envelop (dashed red line) defined by $R_0\pm C(\Omega)/2$  (Eq. \ref{contrastLoss}).

We can further analyze the data for a given tilt $\theta$ of the experiment. The interference pattern becomes clearly visible and the data shows a very good agreement with our model, with no adjustable parameter (Fig. \ref{ratioVsRotationRate}, right). The periodicity of the fringes is determined by the initial transverse velocity and therefore, in the present case, by the tilt angle, as expected for an atomic gyroscope. This shows that our model is an accurate description of the atom interferometer in the presence of rotations, both for contrast loss and the phase terms, within the parameters of our experiment. Moreover, the direct correlation of the atomic signal with the output of the fibre-optic gyroscope proves we are able to extract the rotation signal from a single atomic measurement, while retrieving the acceleration signal of the hybrid accelerometer working in a closed-loop mode (see \cite{Templier2022} and Methods).

In this rigid mode and for our atomic temperature and interrogation time, the operation of the atom interferometer is limited to rotation rates up to $\SI{85}{\milli\radian\per\second}$ ($\SI{5}{\degree\per\second}$) approximately because of the contrast loss. To extend the range of accessible rotation rates and increase the interrogation time, we implement a real-time compensation consisting in maintaining the pointing direction of the reference mirror constant to preserve the overlap of the wave packets and therefore the contrast of the interferometer.

\subsection*{Extension of the dynamic range to higher rotation rates}%%%%%%%%%%%%%%%%%%%%%%%%%%%%%%%%%%%%%%%%%%%%%%%%%%%%%
In order to perform atom interferometry at higher rotation rates, the orientation of the reference mirror in the laboratory frame is kept constant for the duration of the interferometer to maintain the contrast. To that end, we use the signals from the FOGs to control a piezoelectric tip-tilt platform on which the reference mirror is mounted and perform a rotation of the mirror of equal amplitude and opposite direction to the rotation of the device (inertial pointing mode in Fig. \ref{fig:manip}).

\begin{figure}
    \centering
    \begin{subfigure}{\textwidth}
        \includegraphics[width=\textwidth]{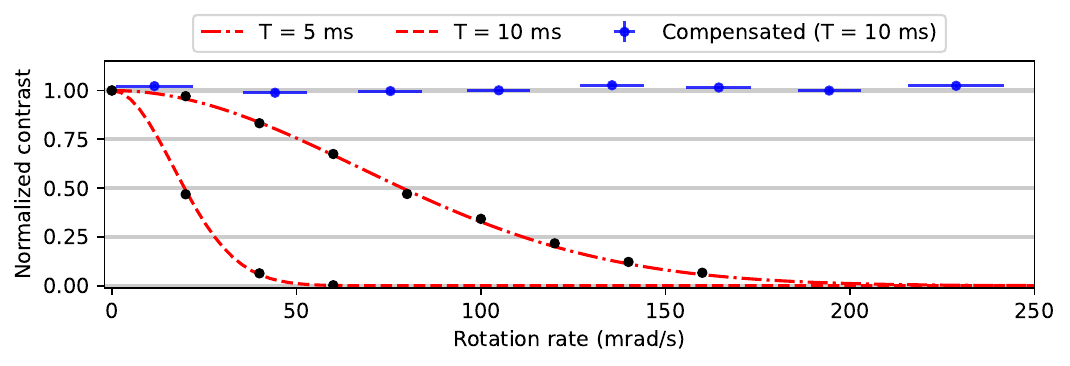}
    \end{subfigure}
    \begin{subfigure}{\textwidth}
        \includegraphics[width=\textwidth]{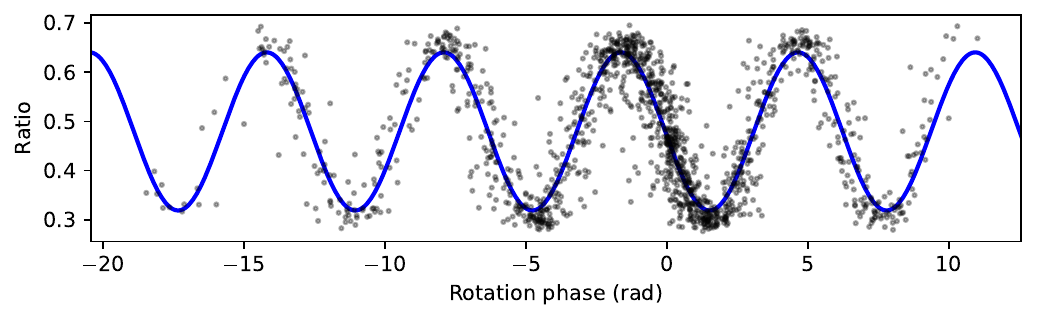}
    \end{subfigure}
    \caption{\textbf{Rotating atom interferometer in the inertial pointing mode}. \textbf{Top:} Normalized contrast $C(\Omega)/C_0$ ($C_0$ is the contrast for a static interferometer) of interference fringes as a function of the rotation rate during the interferometer. The red lines give the contrast calculated with equation \ref{contrastLoss} for a temperature of $\SI{4}{\micro\kelvin}$. The black dots represent experimental measurements of the contrast with a constant rotation of the reference mirror for $T = 5$ and $\SI{10}{\milli\second}$ and a static device. The blue points show the measured contrast when the rotation compensation is active for $2T=\SI{20}{\milli\second}$. Because the rotation rate of the device cannot be precisely controlled with our manual rotation platform, each point correspond to a range of rotation rates. The large horizontal error bars on these points represent the standard deviation of the rotation rate on these ranges. \textbf{Bottom:} Population ratio $R$ as a function of the rotation phase, calculated with equation \ref{phi_comp} from the experimental parameters with $\theta \in [0,30] \degree$, $\Omega \in [-250,250]$ $\SI{}{\milli\radian\per\second}$ and a correction factor $\varepsilon = -0.013$ on the rotation rate of the mirror (see text). The blue curve is a sinusoidal fit to the data points.} 
    \label{fringeWithComp}
\end{figure}

This real-time, active compensation of the rotation consists in applying two angle steps to the tip-tilt platform after the first and second Raman pulses respectively in order to keep the orientation of the mirror constant in the inertial frame. After the first pulse, a measurement of the rotation rate is performed by the gyroscopes and sent to a field-programmable gate array (FPGA) that calculates the angle step the mirror should make before the second pulse to cancel this rotation. The calculated setpoint is then sent to the tip-tilt platform to adjust its position. The same procedure is repeated after the second Raman pulse to adjust the position of the mirror before the third and last Raman pulse such that the orientation of the mirror in the laboratory frame remains the same when the laser pulses are applied. This method of compensation is based on the assumption that the rotation rate of the device is constant between two pulses of the interferometers. Figure \ref{fringeWithComp} (top) shows that it is very effective to maintain the fringes' contrast for an interrogation time $2T=\SI{20}{\milli\second}$ and angular velocities up to, at least, $\SI{250}{\milli\radian\per\second}$ ($\SI{14}{\degree\per\second}$).

The rotation of the reference mirror effectively addresses the drop in the visibility of the atomic fringes but also strongly reduces the phase shift associated with the Coriolis acceleration. However, it also leads to additional phase shifts related to the angular velocity of the mirror in the rotating frame of the device, denoted as $\bm{\Omega_m}$.

With the same assumptions as for the rigid mode, the phase shift due to the rotations, which depends on the rotation rates $\Omega$ and $\Omega_m$ and the tilt $\theta$, is:
\begin{align}
    \begin{split}
        \phi^{inertial}_r \left(\Omega, \Omega_m, \theta\right) = k_{\text{eff}}T^2 &\Bigg[- 2g\sin(\theta)(T+T_{\text{TOF}})(\Omega+\Omega_m) - g\sin(\theta)\Omega_m T\\
        & + z_0^{\text{CA}}\Omega^2 - \left(2 z_0^{\text{MA}} + d_m + 3T\left(g\cos(\theta)T_{\text{TOF}}-\frac{\hbar k_{\text{eff}}}{m}\right)\right)\Omega_m^2\\
        &- 3g\cos(\theta)T_{\text{TOF}}. T(\Omega+\Omega_m)^2\Bigg]
    \end{split}
    \label{phi_comp}
\end{align}
where we have introduced $z_0^{\text{MA}}$, the distance between the atomic cloud and the reference mirror at the beginning of the interferometer and $d_m$, the distance between the mirror's centre of rotation and the surface of the mirror. Note that the first term in this expression, which corresponds to the Coriolis acceleration due to the residual rotation of the mirror in the inertial frame, is close to zero as $\Omega_m \approx -\Omega$. The term $-g\sin(\theta)\Omega_m T$ accounts for the changes in direction and norm of the effective wavevector; the latter being a consequence of an overlap mismatch between the incident and reflected beams. The two terms in $\Omega^2$ and $\Omega_m^2$ of equation \ref{phi_comp} correspond to the centrifugal accelerations linked to the rotations of the device and of the mirror. The last term arises from the composition of motion.

To assess the validity of our rotation phase model in inertial pointing mode, 1400 shots were performed for an interrogation time $2T = \SI{20}{\milli\second}$, rotation rates up to $\SI{250}{\milli\radian\per\second}$ (approximately $\SI{14}{\degree\per\second}$) and tilts between $\SI{0}{\degree}$ and $\SI{30}{\degree}$. The atomic interference pattern is reconstructed by plotting the population ratio as a function of the rotation-induced phase $\phi^{inertial}_r$ calculated from equation \ref{phi_comp} (Fig. \ref{fringeWithComp}, bottom).

Ideally, the angular velocity of the mirror should perfectly compensate the rotation of the device: $\Omega_m=-\Omega$. However, because we have no direct measurement of the mirror's effective angular velocity $\Omega_m$, we introduce a correction factor in our model to take into account a possible discrepancy between the desired and effective rotation rates, and make the substitution $\Omega_m \rightarrow (1+\varepsilon)\Omega_m$ in equation \ref{phi_comp} (see Methods). An optimal reconstruction is found for $\varepsilon = -0.013$. The signal-to-noise ratio (SNR) of the reconstructed fringes, defined as the ratio of the fringes' contrast over the standard deviation of the fit residuals, is 5.3 which corresponds to an acceleration sensitivity $2/(g\cdot \text{SNR}\cdot k_{\text{eff}}T^2) = \SI{24}{\micro\g}$ per shot. For comparison, the signal-to-noise ratio in static conditions is between 20 and 30 over the full angle range $[0, 30]\degree$, for the same interrogation time. We believe that this drop in SNR is due to a degradation of the vibration correction, coming from the excitation of vibration modes of the platform when the device is rotating. This effect can be reproduced by exciting the vibration modes in a non rotating operation (see Methods).

\newpage

\section*{Discussion}
We have demonstrated the operation of an atom interferometer over a large range of angles $([0,30]\SI{}{\degree})$ and rotation rates up to $\SI{14}{\degree\per\second}$, in a regime compliant with operational utilization aboard many platforms (undersea, sea, air and space). The hybridization of the atom interferometer with mechanical accelerometers and optical gyroscopes combined with an accurate model of the expected phase shift allows us to untangle the acceleration and rotation contributions. For strong rotation rates, we implemented a real time compensation of the reference mirror's orientation using a tip-tilt platform and the signal provided by the gyroscopes, thus expanding the range where the contrast of the atom interferometer is maintained. A sensitivity to acceleration of $24$ $\upmu$g can be extracted from the reconstructed fringes, for a total interrogation time of $2T = \SI{20}{\milli\second}$. 

These results and our ability to reconstruct the full acceleration vector for arbitrary orientations in a quasi-static configuration \cite{Templier2022}, show we have the tools necessary to operate quantum inertial sensors based on cold atoms in highly dynamic environments and pave the way for high-performance onboard measurements.

To reach our long term goal of a full quantum ``strapdown" three-axis accelerometer, our rotation compensation system can be extended to a triad including three tip-tilt mirrors and three gyroscopes. The high bandwidth and continuity of the measurement is ensured by the hybridization of the classical sensors and will be pushed forward. Future studies will look into the long-term stability of the three-axis sensor, as well as the systematic errors. The full validation of this strapdown strategy can lead to practical application such as the improvement of the gravity maps, \textit{in situ} calibration of inertial navigation systems or spatial geodesy.

Additionally, our experiment paves the way towards the first onboard atomic gyroscope, working for any angle. The correlation of the phase shift due to rotation with a fibre-optic gyroscope is a first step towards a high dynamic, continuous quantum gyroscope. This is a true alternative to thermal atomic beams \cite{Rakholia2014,Zhao2021} and point source interferometers \cite{Hoth2016}.

%==================================  METHODS =================================================
\newpage
\section*{Methods}
Here, we detail the real time compensation of the acceleration in the closed-loop mode and the limits of the fringe reconstruction due to mechanical modes excitation when the device is rotating. We also present the derivation of the rotation induced phase shift, both for the rigid and inertial pointing modes. We then explain our method to determine the lever arms, and finally, we discuss the correction $\varepsilon$ of the phase shift in the inertial pointing mode.

\subsection*{Closed-loop mode: real-time compensation of vibrations and Doppler effect}
\subsubsection*{General principle}
We have developed a closed-loop operation of our hybrid atom interferometer, using the classical accelerometers to extend the dynamic range and bandwidth of the quantum sensor, while benefiting from the latter's intrinsic high long-term stability. We recall the main ideas for the understanding of the present article. Details can be found in \cite{Templier2022}.

To keep the Raman laser on resonance as the atoms fall, it is necessary to compensate for the Doppler frequency shift caused by the acceleration of the atoms relative to the reference mirror. This shift depends on the orientation of the laser beam with respect to the vertical direction. We use the classical accelerometers installed on the device to determine its orientation by measuring the projections of the gravity vector, and apply an appropriate correction on the laser phase and frequency for any orientation of the device thanks to a FPGA based, real-time feedback system.

We also use the signal of the classical accelerometer installed along the $z$ axis to record the vibrations of the reference mirror, and apply a corresponding correction to the laser phase before the final Raman pulse of the interferometer \cite{Lautier2014}.

Combining these two techniques, we operate the hybrid accelerometer in a closed-loop configuration where the comparison of the classical and quantum acceleration measurements allows us to isolate and correct for the intrinsic bias of the classical device.

Owing to this real-time phase compensation technique, the vibrations and change in orientation do not scan the atomic fringes and the results presented in this paper include only rotation-induced phase shifts.

\subsubsection*{Limits of the fringe reconstruction due to strong vibrations}

\begin{figure}
    \centering
    \includegraphics[width = \textwidth]{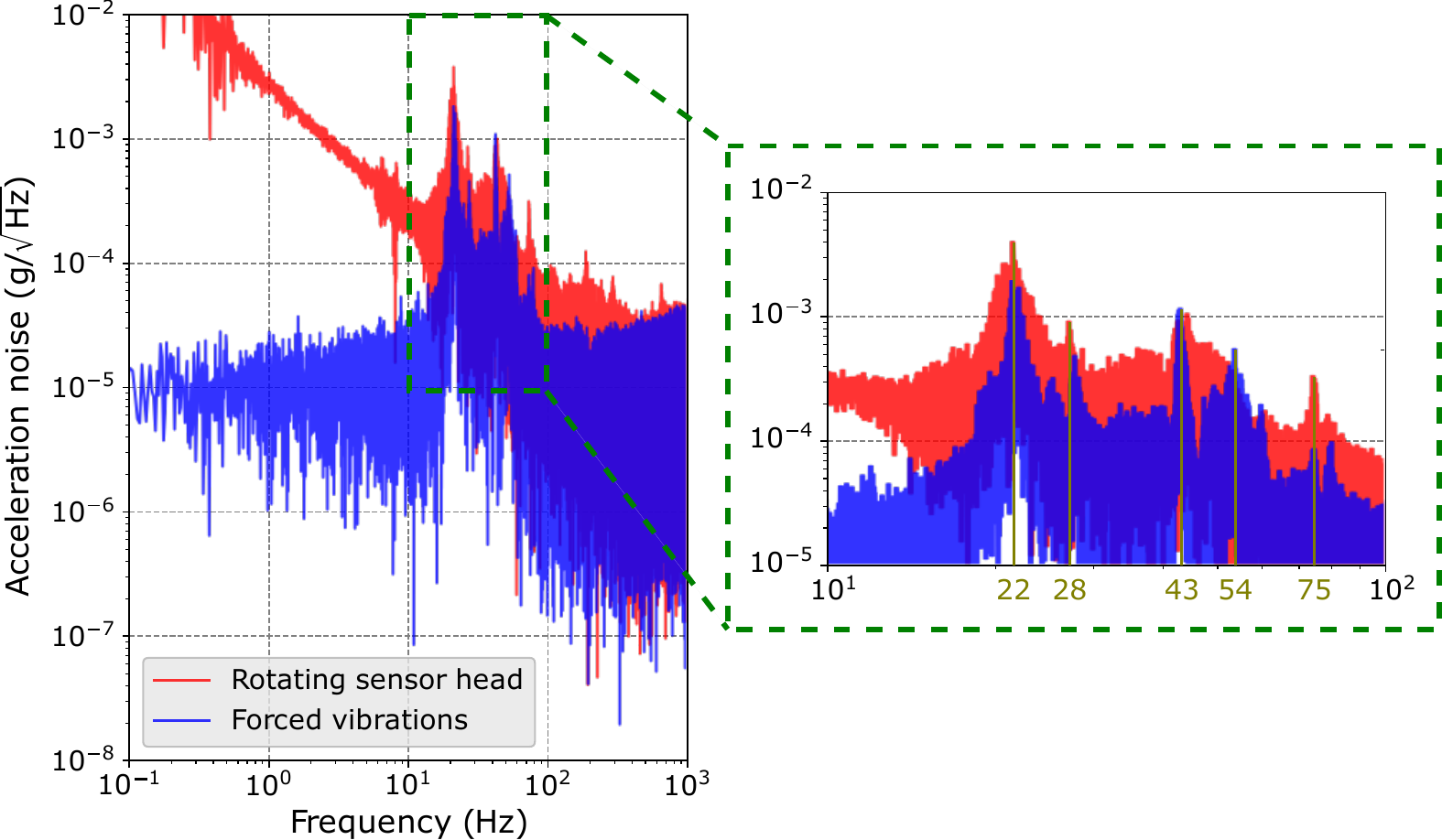}
    \caption{\textbf{Vibration noise on the experiment}. Power Spectrum density of the $z$-axis classical accelerometer signal under rotations of the platform at $|\Omega| \leqslant \SI{250}{\milli\radian\per\second}$ (red) and under forced vibrations produced by tapping the device to excite its vibration modes (blue). Inset: expanded view of the $10 \-- 100$ Hz range where vibrations of the reference mirror are the most critical for the atom interferometer. Brown vertical lines exhibit the highest-amplitude resonances of the mechanical structure. Apart from the peak located at $\SI{54}{\hertz}$, the blue spectrum remains consistently below the noise level obtained under rotations, indicating that the performances measured under forced vibrations are an upper bound on the signal-to-noise ratio which can be reached with such noise, regardless of the device's rotations.} 
    \label{fig:ForcedVibrations}
\end{figure}

The efficiency of the real-time vibration compensation scheme is limited by the performances of the classical accelerometer\footnote{Thales MICAL} attached to the back of the reference mirror, and in particular its bandwidth and self noise, for which we do not have precise characterizations. Our real-time compensation technique also has a limited signal processing capability, meaning that we are not able to compensate for variations in the response function of the accelerometer over the frequency spectrum of interest \cite{Lautier2014}. We simply assume that the amplitude response is constant, and we apply a global delay of $\SI{610}{\micro\second}$ to the acceleration to compensate for the mean phase of the system. This method has proven to be efficient when the vibration spectrum is either close to white noise or centred around a narrow frequency band because the delay corresponds to a perfect phase compensation for a single given frequency \cite{Menoret2018}. Furthermore, in our current setup, the friction generated by the motion of the device and counterweight on the rotation axis of the rotation platform produces high amplitude resonances, especially in the $10 \-- 100$ Hz range as can be seen on the red curve in figure \ref{fig:ForcedVibrations}. Because there are several distinct resonances, or compensation method leads to an imperfect rejection of vibrations over this frequency range, and a significant reduction in the signal-to-noise ratio of the interference fringes and acceleration sensitivity.

In order to estimate the effect of these rotation-induced vibrations, we attempted to reproduce the acceleration noise spectrum obtained in the presence of rotations with a static device. We gently tap the rotation platform to excite its resonance modes while recording atom interference fringes. Under these experimental conditions, represented by the blue curve in figure \ref{fig:ForcedVibrations}, the signal-to-noise ratio of the atomic fringes dropped to 6.2 with a fitted contrast of 27\%. In comparison, the reconstruction of the interference pattern obtained under strong rotations in inertial pointing configuration yielded a signal-to-noise ratio of 5.3 and a contrast of $32\%$. The similarity of these results suggests that our performances are strongly limited by the vibration noise, even when the instrument is rotating. 

\subsection*{Theoretical model of the rotation-induced phase shifts}

In the case of a potential at most quadratic in position and momentum, the phase shift accumulated between the two arms of the atomic interferometer A and B can be described by the mid-point theorem \cite{Antoine2003a, Overstreet2021}. In a Mach-Zehnder geometry, this phase shift simplifies to the expression
\begin{align}
    \phi_{MZ}=\sum_{i=1}^{3} (\bm{k}_{B,i} - \bm{k}_{A,i})\overline{\bm{r}_i}
\end{align}
with $i$ the index of the atom-light interaction, $\bm{k_{A,i}}$ (respectively $\bm{k_{B,i}}$) the wavevector of the laser along the arm A (respectively B) of the interferometer and 
\begin{align}
    \overline{\bm{r}_i} = \frac{\bm{r}_{A,i} + \bm{r}_{B,i}}{2}
\end{align}\\
the position of the centre of mass of the two wavepackets at the $i^{th}$ laser pulse. The full phase shift of the atom interferometer undergoing random accelerations and rotations is estimated by calculating the position of the atoms with respect to the reference mirror at the instant of the laser pulses while taking into account the time evolution of the effective wavevector's norm and orientation. To this end, we carry out calculations in the mobile frame of the device.

In our calculations, we neglect Earth's rotation and assume that the atoms only fall under the action of gravity, leading to the Lagrangian of a free-falling atom in the laboratory frame:
\begin{align}
    \mathcal{L}_0=\frac{1}{2}m\bm{v_0}^2(t)+m\bm{g}.\bm{r_0}(t)
\end{align}
with $\bm{v_0}$ and $\bm{r_0}$ the velocity and position of the atom in the laboratory frame. We then operate a velocity transformation to write the Lagrangian in the mobile frame:
\begin{equation}
    \bm{v_0} = \bm{v_M} + \bm{\Omega_{M}}(t)\cross\bm{r_M}(t) \Rightarrow \mathcal{L}_M=\frac{1}{2}m(\bm{v_M}(t)+\bm{\Omega_{M}}(t)\cross\bm{r_M}(t))^2+m\bm{a}(t).\bm{r_0}(t)
\end{equation}
with $\bm{\Omega_{M}}(t)$ the rotation rate of the device (simply written $\bm{\Omega}(t)$ below to simplify notations) and $\bm{a}(t)$ the linear acceleration of the atoms relative to the device which can be decomposed as $\bm{a}(t) = \bm{g} - \bm{a_{M}}(t)$. The term $\bm{a_M}(t)$ includes both the DC component of the sensor's acceleration $\mathbf{a_{s}}(t)$ and vibrations $\bm{a_{vib}}(t)$. $\bm{v_M}(t)$ and $\bm{r_M}(t)$ are the velocity and position of the atoms in the mobile frame. Taking the origin of this frame at the mobile's centre of rotation, which belongs to both inertial frames, all velocities and positions are defined with respect to this point and we can write $\bm{r}(t) := \bm{r_0}(t) = \bm{r_M}(t)$. We have:
\begin{align}
    \mathcal{L}&=\frac{1}{2}m(\bm{\dot r}(t)+\bm{\Omega}(t)\cross\bm{r}(t))^2+m\bm{a}(t).\bm{r}(t) \\
    &=\frac{1}{2}m\left[(\bm{\dot r}^2(t)+2\bm{\dot r}(t).\left(\bm{\Omega}(t)\cross\bm{r}(t)\right) + \left(\bm{\Omega}(t)\cross\bm{r}(t)\right)^2\right]+m\bm{a}(t).\bm{r}(t)
\end{align}
Calculating the derivatives of the Lagrangian with respect to the atom's position and velocity and using the Euler-Lagrange equation leads to the equation of motion:
\begin{equation}
    \bm{\ddot r}(t) + 2\left(\bm{\Omega}(t)\cross\bm{\dot r}(t)\right) + \bm{\Omega}(t)\cross\left(\bm{\Omega}(t)\cross\bm{r}(t)\right) + \bm{\dot \Omega}(t)\cross\bm{r}(t) = \bm{a}(t)
\end{equation}
where we recognize on the left-hand side in order of appearance the relative, Coriolis, centrifugal and tangential accelerations, the latter two forming what is called the driving acceleration. To solve this equation and calculate the atomic trajectories we use a polynomial decomposition of the atoms' position $(\bm{r}(t)=\sum_{j\leq 3} c_j\times t^j)$, considering, at first, constant accelerations and rotation rates and truncating to third order terms. Ultimately, we will consider in the following only rotations with axes in the plane transverse to the effective wavevector of the laser $\bm{\Omega}=\left(\Omega_x, \Omega_y, 0\right)^\text{T}$.

All the inertial components, namely acceleration, velocity and position of the atomic wave packet relative to the reference mirror are defined with respect to their values at the instant of the first laser pulse (taken as initial conditions). Acceleration is considered constant, and velocity can be computed from the moment the atoms are released from the optical molasses, since the atoms are trapped and have no relative velocity in the frame of the experimental chamber prior to this point.

Then, as mentioned above, calculating the phase delay accumulated between the two arms A and B of the interferometer can be achieved by using the mid-point theorem written as 
\begin{align}
    \begin{split}
        \phi_{at} &= \bm{k_{\text{eff}}}(0).\left(\bm{r_{at}}(0)-\bm{r_m}(0)\right) - 2\bm{k_{\text{eff}}}(T).\left(\frac{\bm{r_{at}^A}(T)+\bm{r_{at}^B}(T)}{2}-\bm{r_m}(T)\right) \\
        &+\bm{k_{\text{eff}}}(2T).\left(\frac{\bm{r_{at}^A}(2T)+\bm{r_{at}^B}(2T)}{2}-\bm{r_m}(2T)\right)  
    \end{split}
    \label{eq: MidPoint}
\end{align}
with $\bm{k_{\text{eff}}}(t)$, $\bm{r_{at}}(t)$ and $\bm{r_m}(t)$ the effective wavevector of the laser, the position of the atoms and the position of the mirror at the instant $t$ respectively. The superscripts $A,B$ on the atomic cloud's position vectors denote the interferometer's path they follow.

\subsubsection*{Phase shift in rigid mode}

Assuming the reference mirror is immobile in the device's reference frame such that the effective wavevector's norm and orientation remain constant in that frame and, setting $\mathbf{k}(t)=k_{\text{eff}}.\mathbf{u_z}$ so that only rotations on the $x$ and $y$ axes will appear, the phase shift at the output of the interferometer calculated using the mid-point theorem is:
\begin{align}
    \begin{split}
        \phi_{at}  = k_{\text{eff}}T^2 &\big[ a_z + 2(v_x + a_x T)\Omega_y - 2(v_y + a_y T)\Omega_x \\
        &+ \left(z_0 - 3v_z T + 2x_0\Omega_y T - 2y_0\Omega_x T\right) \Omega^2 \big]
    \end{split}
    \label{atomic_phase}
\end{align}
with $\Omega^2 = \Omega_x^2 + \Omega_y^2$ and under the assumption of infinitely short laser pulses. Here, we recognize the relative acceleration, two Coriolis terms, and a last term corresponding to the centrifugal acceleration.

Due to our hybridization scheme (see Methods and \cite{Templier2022}), the measurement of the classical accelerometer attached to the reference mirror is subtracted in real time from the atomic acceleration. To account for this, one can define an equivalent phase shift from the classical acceleration using the formalism of the sensitivity function \cite{Cheinet2008}. This method yields a phase shift equal to
\begin{align}
    \begin{split}
        \phi_{cl} &= k_{\text{eff}}T^2\left[ a_z + (z_0^{cl} + v_z^{cl}T)\Omega^2 \right]\\
        &= k_{\text{eff}}T^2\left[ a_z + z_0^{cl}\Omega^2 \right]
    \end{split}
    \label{eq:ClassicalPhase}
\end{align}
considering that the classical sensor, fixed in the experimental chamber's frame, has no relative velocity $v_z^{cl}$. The subtraction of the two expressions in equations \ref{atomic_phase} and \ref{eq:ClassicalPhase} finally gives us the inertial phase shift remaining to be compensated for:
\begin{align}
    \begin{split}
        \phi_r^{rigid}  = k_{\text{eff}}T^2 &\big[ 2(v_x + a_x T)\Omega_y - 2(v_y + a_y T)\Omega_x \\
        &+ \left(z_0^{\text{CA}} - 3v_z T + 2x_0\Omega_y T - 2y_0\Omega_x T\right) \Omega^2 \big]
    \end{split}
    \label{phi_rigid_full}
\end{align}
where $z_0^{\text{CA}}$ denotes the initial distance between the atomic wave packet and the classical accelerometer along the $z$ axis.\\

\subsubsection*{Phase shift in inertial pointing mode}

Here we calculate the phase shift for the inertial pointing mode, where we apply a rotation to the mirror in order to keep the orientation of the Raman wavevector constant in the frame of the laboratory. The contributions to the atomic phase shift are summarized in Table \ref{tab:phi_init_at} where we have introduced $\Omega_m^2 = \Omega_{mx}^2 + \Omega_{my}^2$ the norm of the mirror's rotation rate and $d_{\text{COR}}$ the distance between the centre of rotation of the chamber and of the mirror, respectively. The total phase shift is the sum of all these contributions. Note that the rotation of the mirror cancels out the Coriolis effect almost entirely ($\Omega_i\simeq -\Omega_{mi}$) and is at the origin of additional terms in the centrifugal acceleration.
\renewcommand{\arraystretch}{1.5}
\begin{table}[h]
    \centering
    \begin{tabular}{|l|l|}
    \hline
        \textbf{Acceleration term} & \textbf{Phase shift expression}\\ 
        \hline
        \hline
        Relative & $k_{\text{eff}}T^2 a_z$\\
        \hline
        Mirror's rotation & $k_{\text{eff}}T^3(a_x \Omega_{my} - a_y \Omega_{mx})$\\
        \hline
        Residual Coriolis & $2k_{\text{eff}}T^2\big[(v_x + a_x T)(\Omega_y + \Omega_{my}) - (v_y + a_y T)(\Omega_x + \Omega_{mx})\big]$\\
        \hline
        Centrifugal (chamber) & $k_{\text{eff}}T^2(z_0 + 2 x_0 \Omega_y T - 2 y_0 \Omega_x T)\Omega^2$\\
        \hline
        \multirow{2}{4em}{Centrifugal (mirror)} & $-k_{\text{eff}}T^2(2 z_0 - 2 d_{\text{COR}} - d_m + 3 v_z T + 3 v_{rec} T $\\
        &$+ 3 x_0 \Omega_{my} T - 3 y_0 \Omega_{mx} T)\Omega_m^2$\\
        \hline
        \multirow{2}{4em}{Others}& $3k_{\text{eff}}T^3\big[(x_0\Omega_y - y_0\Omega_x)(\Omega_x\Omega_{mx} - \Omega_y\Omega_{my})$ \\
        & $ - 3v_z(\Omega_x^2 + 2\Omega_x\Omega_{mx} + \Omega_{mx}^2 + \Omega_y^2 + 2\Omega_y\Omega_{my} + \Omega_{my}^2)\big]$\\
        \hline
    \end{tabular}
    \caption{Contributions to the atomic phase shift of the atom interferometer in the inertial pointing mode, up to the third order in $T$.}
    \label{tab:phi_init_at}
\end{table}

Since the mechanical accelerometer is not sensitive to the rotation of the mirror, the terms subtracted from the previous expression are the same as in equation \ref{eq:ClassicalPhase}. To simplify notations, we will write:
\begin{itemize}
    \item $z_0^{\text{CA}} = z_0 - z_0^{cl}$ the distance between the classical sensor and the atomic wave packet at the instant of the first pulse,
    \item $z_0^{\text{MA}} = z_0 - d_{\text{COR}} - d_m$ the distance between the atomic cloud and the reference mirror's surface at the same time,
    \item $\delta\Omega_i=\Omega_i+\Omega_{mi}$ ($i=x,y$) the residual rotation rate on a given axis arising from an imperfect compensation.
\end{itemize} 
The expression of the phase shift in the the closed-loop mode with the classical sensor becomes:
\begin{align}
    \begin{split}
        \phi^{inertial}_r  = k_{\text{eff}}T^2 &\big[ 2(v_x + a_x T)\delta\Omega_y - 2(v_y + a_y T)\delta\Omega_x + (a_x \Omega_{my} - a_y \Omega_{mx})T \\
        &+ (z_0^{\text{CA}} + 2 x_0 \Omega_y T - 2 y_0 \Omega_x T)\Omega^2 - 3v_zT(\delta\Omega_x^2 + \delta\Omega_y^2) \\
        &- (2 z_0^{\text{MA}} + d_m + 3T(v_z + v_{rec} + x_0\Omega_{my} - y_0\Omega_{mx}))\Omega_m^2 \\
        &+ 3T(x_0\Omega_y - y_0\Omega_x)(\Omega_x\Omega_{mx} - \Omega_y\Omega_{my}) \big]
    \end{split}
    \label{phi_init_full}
\end{align}

\subsubsection*{Tangential acceleration}

The derivations of the phase in the rigid (Eq. \ref{phi_rigid_full}) and inertial (Eq. \ref{phi_init_full}) modes were made with the assumption that the angular velocity of the device is constant. If we now consider that the angular velocity varies at most linearly with time, a tangential acceleration of the atoms, related to the angular acceleration $\dot{\Omega}$, will result in an additional phase shift:
\begin{align}
    \phi_{tang} = k_{\text{eff}}T^2\left(-\dot{\Omega}_x\left(y_0+v_yT\right)+\dot{\Omega}_y\left(x_0+v_xT\right)\right)
\end{align}
The instantaneous angular acceleration can be estimated by calculating the gradient between two successive measurements of the fibre-optic gyroscopes. An average value over the duration of the interferometer can then be used to calculate the phase term associated with the tangential acceleration. For the data set in inertial pointing mode, the measured angular accelerations follow a Gaussian distribution centred in zero with a standard deviation of $\SI{72}{\milli\radian\per\second\squared}$. We find that, in our experiments, this term has a small influence on the overall rotation-induced phase shift --- the phase shift associated with the angular acceleration is one or two orders of magnitude below the total phase shift --- and produces no significant difference in the analysis of the data.

For more general cases, with a significant time dependence of the rotation rate  during the atom interferometer, it is possible to use a temporal response function to model the expected phase shift \cite{struckmann2023platform}. Our goal is to develop a method which can be implemented in real time, with our FPGA for instance. The simple approach of our article is easier to implement than the temporal response function which is consequently beyond the scope of this work.

\subsection*{Determination of the lever arms}

The centrifugal accelerations in the phase shifts of equations \ref{phi_rigid_full} and \ref{phi_init_full} require evaluating different lever arms on the apparatus, namely the distances $d_m$, $z_0^{\text{MA}}$ and $z_0^{\text{CA}}$.

The distance $d_m$ between the surface of the mirror and its centre of rotation is a property of the tip-tilt device and is given by its datasheet ($d_m = \SI{7}{\milli\meter})$.\\
To determine the position of the cloud with respect to the mirror at the beginning of the interferometer ($z_0^{\text{MA}})$, we apply a frequency jump $\Delta f$ to the laser at the $\pi$ pulse of a Mach-Zehnder interferometer (without rotations) which introduces an additional phase shift equal to \cite{Xu2021}:
\begin{align}
    \Delta \phi = \frac{8 \pi d_{\pi}}{c} \Delta f
\end{align}
with $d_{\pi}$ the distance between the atomic cloud and the reference mirror at the instant of the $\pi$ pulse and $c$ the speed of light. The distance $d_\pi$ is extracted from the measured phase shift as a function of the applied frequency shift. The position $z_0^{\text{MOT}}$ of the atomic cloud when released from the MOT is obtained by adding $d_{\pi}$ to the distance traveled by the atoms up to the second pulse of the interferometer. This operation is repeated for different frequency jumps and times of flight to find a reference position: $z_0^{\text{MOT}} = \SI{128(4)}{\milli\meter}$.

The distance $z_0^{\text{MA}}$ is then calculated for each data point by summing $z_0^{\text{MOT}}$ and the distance traveled before the first interferometer pulse, taking into account the time of flight of the atoms before the interferometer and the tilt angle of the device $z_0^{\text{MA}} = z_0^{\text{MOT}} + \frac{1}{2}gT_{\text{TOF}}^2\cos\theta$.

Finally, the distance between the atoms and the proof mass of the classical accelerometer is calculated by adding $z_0^{\text{MA}}$ to the distance between the accelerometer and the mirror which is known from the mechanical drawings of the experiment and the datasheet of the accelerometer: $z_0^{\text{CA}} = z_0^{\text{MA}} + \SI{96(1)}{\milli\meter}$.

\subsection*{Correction of the phase shift in the inertial pointing mode}

\begin{figure}
    \centering
    \includegraphics[width = \textwidth]{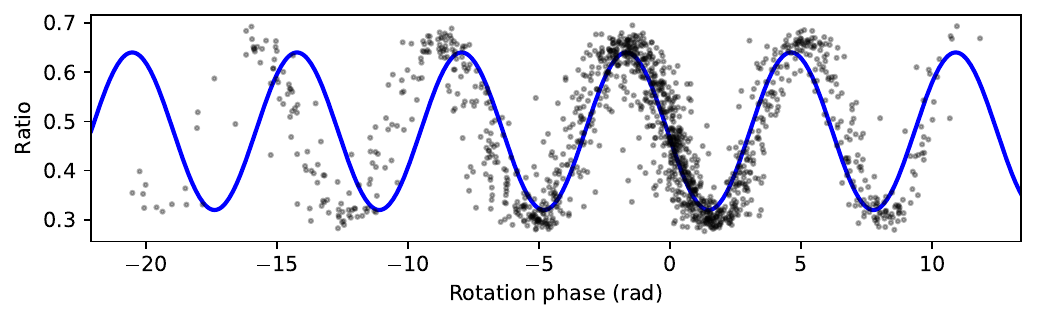}
    \caption{\textbf{Raw correlation between the calculated phase shift and the atom interferometer in the inertial mode}. Population ratio $R$ as a function of the rotation phase, calculated with equation \ref{phi_comp} from the experimental parameters with $\theta \in [0,30] \degree$, $\Omega \in [-250,250]$ $\SI{}{\milli\radian\per\second}$. The blue curve is a $2\pi$-periodic sinusoidal function which corresponds to the theoretically expected fringes.} 
    \label{residuals}
\end{figure}

The calculation of the rotation phase in the inertial pointing mode reveals an anomaly in the reconstructed interference pattern. Figure \ref{residuals} shows the population ratio measured at the output of the interferometer as a function of the rotation phase calculated with equation \ref{phi_comp} when no correction is performed. This interferogram clearly exhibits a sinusoidal pattern but not with the expected $2\pi$-periodicity. This shows that the evaluation of the phase shift is not perfect even if the periodicity of the reconstructed fringes indicates a strong correlation between the true and estimated phase shifts. We therefore look at the potential systematic errors in the determination of the experimental parameters: the tilt angle $\theta$, the device's rotation rate $\Omega$ and the mirror's rotation rate $\Omega_m$.

For each of them, we evaluate the error that would explain our signal, and rule out the ones which are incompatible with independent estimations. We find that all the suspected effects can be rejected, except for a calibration error of the tip-tilt mirror.

\subsubsection*{Tilt angle}
The tilt $\theta$ of the experiment with respect to gravity determines the transverse velocity of the atoms and intervenes in several terms of the phase shift. This angle is calculated from the measurements of the classical accelerometers of the device. To evaluate the effect of an error in the measurement of the tilt angle, we make the substitution $\theta\rightarrow (1+\varepsilon)\theta$ in the expression of the phase shift. For each value of $\varepsilon$, the corrected data is fitted with a sinusoidal function for which the contrast and offset are fixed, while the period and phase are the fitting parameters. We keep the value $\varepsilon$ leading to a $2\pi$-periodic signal and find $\varepsilon = -0.103$.

We can reject this error on the tilt angle for two reasons. First, the measurement of the angle from the accelerometers is consistent with the goniometer of our rotation platform whereas such a large error (10 \% corresponds to $3\degree$ at $30\degree$) would be clearly visible. Second, this error disagrees with the analysis done in the rigid mode. Introducing a 10 \% error on the tilt angle for the rigid mode leads to a severely degraded reconstruction of the interference pattern.

\subsubsection*{Scale factor of the gyroscopes}
The rotation rate of the device is directly measured by the fibre-optic gyroscopes. An error in the scale factor of the gyroscopes will lead to a systematic error in the calculation of the rotation induced phase shift which could explain our observations on the interferogram. Now performing the substitution $\Omega\rightarrow(1+\varepsilon)\Omega$, we find that a correction $\varepsilon = 0.016$ allows us to retrieve a $2\pi$-periodicity.
Nevertheless, the scale factors of the gyroscopes have been characterized with a $10^{-5}$ relative precision, which is much smaller than the discrepancy we find. We therefore exclude this from the list of possible sources of error.

\subsubsection*{Mismatch of tip-tilt and device axes}

The tip-tilt mirror rotates around two axes which are determined by the position of its piezoelectric actuators. These axes are at an angle $\gamma \approx 45\degree$ with respect to the measurement axes of the fibre-optic gyroscopes on the device. An error on the angle $\gamma$ leads to an imperfect compensation of the device's rotation. This, in turn, produces a non-zero Coriolis phase shift as the motion of the mirror in the inertial frame of reference is not a pure translation. In the general expression of the rotation induced phase (Eq. \ref{phi_init_full}), this corresponds to the term $2(v_x + a_x T)\delta\Omega_y - 2(v_y + a_y T)\delta\Omega_x$.

We thus assume an error on the angle $\gamma$ and calculate the corresponding values of $\delta\Omega_x$ and $\delta\Omega_y$ to reconstruct the interference pattern. We also allow for a non-zero initial velocity along the $x$ direction ($v_x$). Using the same criterion of $2\pi$-periodicity as above, we find an optimum for an error of $\delta \gamma \approx 9\degree$ which is too large a value to be credible. From design and independent measurements we estimate $\delta\gamma = \pm 2\degree$ at most.

\subsubsection*{Scale factor of the tip-tilt mirror}
If the scale factor of the tip-tilt mirror (angular displacement versus set voltage) is not perfectly calibrated, there could be a discrepancy between the desired and effective rotation rate of the mirror that would lead to errors in the calculated phase shift (\Eq \ref{phi_init_full}).

Making the substitution $\Omega_m\rightarrow(1+\varepsilon)\Omega_m$ in the expression of the phase shift, we find an optimal correction of the fringes' periodicity for $\varepsilon = -0.013$.

We have carried out an optical characterization of the mirror by shining a laser on the mirror and measuring the displacement of the reflected beam on a position sensitive photodiode. These measurements are compatible with the expected value of the correction factor $\varepsilon$, with a precision at the 1\% level. A more accurate measurement is required to confirm the calibration error of the tip-tilt mirror. It is, however, our favoured hypothesis to explain the observed anomaly in the reconstructed fringes in the inertial pointing mode.

\newpage
\bibliography{References}
\bibliographystyle{ScienceAdvances}

\section*{Acknowlegments}
This work is supported by the French national agencies ANR (Agence Nationale pour la Recherche) under grant no. ANR-20-CE47-0008 MiniXQuanta  and PEPR (Programmes et Equipements Prioritaires de Recherche) France Relance 2030 QAFCA grant no.  ANR-22-PETQ-0005 QAFCA. We acknowledge support from the Naquidis Center (VEGA project). V. M. acknowledges support from the France Relance initiative.

\section*{Author information}
\subsection*{Contributions}
P.B. and B.B. conceived the project. Q.A.C. and S.T. built the apparatus, Q.A.C., S.T., and R.A. designed and implemented the rotation compensation system, Q.A.C., C.D.C., and B.B. worked out the theoretical model of the phase shift with contributions from V.J. and V.M., Q.A.C., C.D.C., and V.J. performed experiments, Q.A.C., and V.J. carried out the data analysis. B.B. and V.M. supervised the experiments, data analysis and calculations. B.B., B.D., P.B. and V.M. coordinated and administrated the project. B.B., Q.A.C., V.J., and V.M. wrote the manuscript. All authors discussed and reviewed the manuscript.
%==========================================================================

%%%%%%%%%%%%%%%%%%%%%%%%%%%%%%%%%%%%%%%%%%%%%%%%%%%%%%%%%%%%%%%%%%%%%%%%%%%%
\end{document}